\documentclass[aps,pre,twocolumn,nofootinbib,tightenlines,superscriptaddress,nopacs,amsmath,amssymb,final,letterpaper]{revtex4-1}

\usepackage[utf8]{inputenc}
\usepackage{calc}
\usepackage{graphicx}
\usepackage{amsmath,amssymb,amsthm}
\usepackage{bbold}
\usepackage{bm}
\usepackage{color}
\usepackage[dvipsnames]{xcolor}
\usepackage{enumitem}
\usepackage{tikz}
\usepackage{float}
\usepackage[percent]{overpic}

\DeclareMathOperator*{\argmin}{arg\,min}

\newcommand{\pder}[2]{\frac{\partial #1}{\partial #2}}

\newcommand{\abs}[1]{\vert #1\vert}

\def\multiset#1#2{\ensuremath{\left(\kern-.3em\left(\genfrac{}{}{0pt}{}{#1}{#2}\right)\kern-.3em\right)}}

\usepackage[colorlinks=true,linkcolor=blue,urlcolor=blue,citecolor=blue,anchorcolor=blue]{hyperref}

\usepackage{multirow}
\usepackage{dcolumn}% Align table columns on decimal point
\usepackage[table]{xcolor}
\usepackage{booktabs}% for the bottomline in tables

\begin{document}
\title{Hypergraph backboning}

\author{Alec Kirkley}
\email{alec.w.kirkley@gmail.com}
\affiliation{School of Computing and Data Science, University of Hong Kong, Hong Kong SAR, China}

\author{Helcio Felippe}
\affiliation{Department of Network and Data Science, Central European University, Vienna, Austria}

\author{Federico Malizia}
\affiliation{Department of Network and Data Science, Central European University, Vienna, Austria}

\author{Federico Battiston}
\email{battistonf@ceu.edu}
\affiliation{Department of Network and Data Science, Central European University, Vienna, Austria}
\affiliation{Department of AI, Data and Decision Sciences, Luiss University of Rome, Rome, Italy}

\begin{abstract}
Hypergraphs provide a natural framework for describing complex networked systems with higher-order, non-dyadic interactions. Due to their high dimensionality and often redundant structure, a key challenge is to develop methods that simplify hypergraph representations while preserving the essential structure of interactions. Here we present a principled, efficient, and non-parametric information-theoretic method for pruning nested and/or redundant structures in hypergraphs, enabling a minimal representation of higher-order interactions in the presence of local heterogeneity. Our approach naturally extends to weighted hypergraphs, where higher-order topology and hyperedge weights combine to identify the system's structural backbone. We validate the method on controlled synthetic hypergraphs and apply it to empirical datasets from diverse domains, demonstrating substantial sparsification without loss of core structural information.
\end{abstract}

%%%%%%%%%%%%%%%%%%%%%%%%%%%%%%%%%%%%%%%%%%%%%%%%%%%
%%%%%%%%%%%%%%%%%Intro%%%%%%%%%%%%%%%%%%%%%%%%%%%%%
%%%%%%%%%%%%%%%%%%%%%%%%%%%%%%%%%%%%%%%%%%%%%%%%%%%

\maketitle

\section{Introduction}

Many natural and artificial systems, from biological signaling pathways to ecological communities and social contagion processes, exhibit interactions that simultaneously involve more than two constituents~\cite{battiston2020networks,battiston2021physics,bianconi2021higher,bick2023higher}. In such settings, hypergraphs~\cite{berge1984hypergraphs}, where hyperedges encode interactions among an arbitrary
number of nodes, provide a parsimonious framework to represent these non-dyadic, higher-order interactions. Recent methodological advances have enabled the systematic exploration and modeling of higher-order architectures in empirical domains~\cite{benson2018simplicial, petri2018simplicial,
contisciani2022inference, di2024percolation}, uncovering collective behaviours~\cite{battiston2026collective} that are not easily described by pairwise interactions in processes ranging from
contagions~\cite{iacopini2019simplicial, burgio2024triadic,
ferraz2024contagion} and diffusion~\cite{di2024dynamical} to
synchronization~\cite{skardal2019abrupt,millan2020explosive,zhang2023higher} and evolutionary dynamics~\cite{alvarez2021evolutionary,
civilini2024explosive}. Nevertheless, the high dimensionality of real-world hypergraphs makes higher-order analyses more complex and computationally demanding than their pairwise counterparts, posing the problem of constructing simpler representations that preserve the essential structural heterogeneity of the original hypergraph.

Graph sparsification is a well-studied topic in the modern network science landscape, benefiting from a mature toolkit of methods where network edges can be pruned based on spectral properties~\cite{spielman2008graph,spielman2011spectral}, edge positions and weights relative to a statistical null model~\cite{tumminello2011statistically,dianati2016unwinding,casiraghi2017relational,serrano2009extracting,marcaccioli2019polya,kirkley2025fast}, shortest-path structure~\cite{grady2012robust}, and community structure~\cite{rajeh2022modularity}. Comprehensive reviews of such methods are available that provide coverage of additional methods for network sparsification~\cite{neal2022backbone,yassin2023evaluation}. By contrast, methods for higher-order network sparsification remain comparatively scarce. A number of existing works have focused on identifying hyperedges and simplices that are unlikely to exist under a structural null model~\cite{musciotto2021detecting,musciotto2022identifying}, but this requires specification of a significance level which determines the sparsity of the inferred backbone. Other methods aim to extract the key interaction orders contributing to a hypergraph's topological features~\cite{landry2024filtering,ceria2025relevance,kirkley2025structural}, but are unable to capture the structural backbone that exists simultaneously at different orders of interaction. Additionally, methods have been proposed to prune higher-order structure based on its dynamical properties~\cite{zhang2024diffusion,lucas2026reducibility}, though this requires specification of the dynamics or dynamical operator of interest (e.g. the graph Laplacian) and does not necessarily extract the structurally critical higher-order interactions present~\cite{kirkley2025structural}. Finally, all such methods deal exclusively with binary hypergraphs, neglecting the role of interaction strength. In short, a principled, parameter-free framework for constructing unweighted and weighted hypergraph backbones that retain essential structure and key connections at all orders of interactions is currently still lacking. 

The minimum description length (MDL) principle from information theory, which states that the best model for a dataset is the one that permits its shortest transmission (in bits of information) offers a natural foundation for such a sparsification framework~\cite{rissanen1978modeling}. The MDL principle has proven effective for various pairwise graph analyses including community detection~\cite{peixoto2017nonparametric,peixoto2019bayesian,kirkley2022spatial,morel2024bayesian}, network ensemble comparison~\cite{hebert2024network,peixoto2023implicit}, and quantifying the similarity among multiple graphs or their partitions~\cite{coupette2021graph,coupette2022differentially,kirkley2023compressing,felippe2024network,jerdee2024mutual}. Here we utilize the MDL principle to develop a principled, interpretable, and computationally efficient method for identifying the structural backbone and the key interactions present in a higher-order system at all orders. By framing sparsification as a lossless compression problem, our method learns the optimal backbone directly from an observed hypergraph without requiring user-specified significance thresholds or dynamical models, avoiding confirmation biases inherent to parameter-driven approaches. Our framework exploits topological redundancies such as nested hyperedges~\cite{lotito2022higher,larock2023encapsulation,landry2024simpliciality,gallo2024higher}, and is the first method to also naturally accommodate hyperedge weights arising from repeated or noisy measurements, yielding a minimal representation of higher-order interactions by solving a tradeoff between topological features and interaction strength~\cite{kirkley2025fast}. This approach consistently recovers planted structural regularities in synthetic hypergraphs with controlled levels of nestedness and redundancy, and provides substantial sparsification when applied to empirical hypergraphs across domains. Our proposed method enables the identification of the key higher-order interactions that comprise complex systems, providing a principled pathway toward efficient and interpretable higher-order network analyses. 

\begin{figure}[t!]
\includegraphics[width=0.50\textwidth]{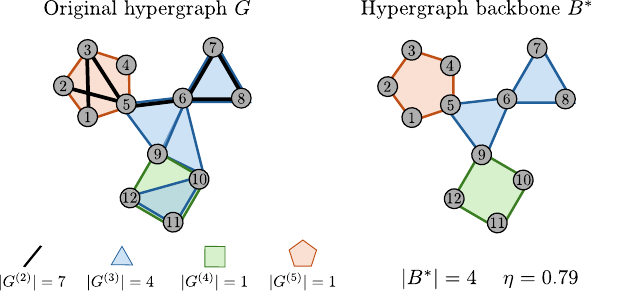}
\caption{
    \textbf{Higher-order backboning of a hypergraph.}
    Original hypergraph~$G$ with interactions ranging from dyads
    (shown on the bottom as thick black edges) and triplets (blue
    triangles) to 4- and 5-body hyperedges (green squares and orange
    pentagons, respectively).  Structural backbone~$B^\ast$ obtained
    through hypergraph backboning by solving Eq.~\eqref{eq:argmin}, along with the corresponding compression ratio (Eq.~\eqref{eq:comp-ratio}).
}
\label{fig:fig1_diagram}
\end{figure}

%%%%%%%%%%%%%%%%%%%%%%%%%%%%%%%%%%%%%%%%%%%%%%%%%%%
%%%%%%%%%%%%%%%%%Methods%%%%%%%%%%%%%%%%%%%%%%%%%%%
%%%%%%%%%%%%%%%%%%%%%%%%%%%%%%%%%%%%%%%%%%%%%%%%%%%

\section{Backboning unweighted hypergraphs}
\label{sec:unweighted}

Let $G$ be a hypergraph on $N$ nodes consisting of undirected hyperedges with no self-edges (i.e., no repeated nodes within a given hyperedge). Also let $L$ be the number of unique hyperedge sizes (orders) in $G$. The goal of our method is to identify a backbone~$B\subseteq G$ consisting of a subset of hyperedges in $G$ that retains its essential structure. We will start by deriving a method for extracting the structural backbone of $G$ when its hyperedges are unweighted (i.e., each hyperedge~$e\in G$ is just a tuple of nodes), focusing on the nested and overlapping hyperedge structures only present in higher-order networks. 

Using the MDL principle, we will construct an objective function~$\mathcal{L}(G,B)$ which assigns a codelength to each subset~$B\subseteq G$ telling us how well $B$ compresses the structure of $G$ when used as a backbone to summarize its key interactions. According to the Kraft inequality from information theory~\cite{mackay2003information}, if $P(x)$ is a properly normalized probability distribution over elements~$x\in \mathcal{X}$ in a finite space~$\mathcal{X}$, one can construct a valid prefix code in which each element~$x$ has a codelength of $-\log P(x)$ bits. (We will let $\log \equiv \log_2$ for brevity.) The simplest such encoding is a fixed-length code that corresponds to a uniform distribution over $x\in \mathcal{X}$, which assigns each element a codelength~$-\log \frac{1}{\abs{\mathcal{X}}} = \log \abs{\mathcal{X}}$. We can thus construct a sequence of fixed-length codes to derive a parameter-free codelength~$\mathcal{L}(G,B)=\mathcal{L}(B)+\mathcal{L}(G\vert B)$ that represents the information required for a lossless information transmission process in which the hypergraph backbone $B$ is transmitted first using $\mathcal{L}(B)$ bits of information, and $G$ is then transmitted using $\mathcal{L}(G\vert B)$ bits by exploiting its structural regularities captured by $B$. 

Encoding the backbone~$B$ is equivalent to encoding each of its hyperedges~$p$ (which we call ``parent'' hyperedges), which in turn requires encoding the size~$\abs{p}$ and the nodes within $p$ (we will ignore the information required to encode the constants $L$ and $N$, as this is small and independent of the backbone choice~$B$). There are $L$ possibilities for the size~$\abs{p}$, so under a fixed-length code this quantity can be encoded with $\log L$ bits. Similarly, there are ${N\choose \abs{p}}$ possibilities for the nodes that comprise $p$ (not permitting repetitions), so these nodes can be transmitted using $\log {N\choose \abs{p}}$ bits under a fixed-length code. Putting the two information costs together, it costs us
\begin{align}\label{eq:Hent}
H(p) = \log L+\log {N\choose \abs{p}}   
\end{align}
bits to transmit the parent hyperedge~$p\in B$, and summing over the hyperedges in $B$, it costs us
\begin{align}
\mathcal{L}(B) = \sum_{p\in B}H(p)
\end{align}
bits to transmit the backbone~$B$. We can see that Eq.~\eqref{eq:Hent} provides a combinatorial entropy expression for the hyperedge~$p$, and the sum of these entropies gives the total description length~$\mathcal{L}(B)$ for the backbone. 

We can then transmit $G$ by using its backbone~$B$ as a structural template from which the remaining hyperedges~$c\in G\setminus B$ (which we call ``child'' hyperedges) are transmitted. The method exploits dependencies among hyperedges of different orders, such as overlap and nestedness. Hyperedges that can be efficiently encoded from a more informative parent hyperedge are then treated as redundant, and the MDL encoding provides the formal criterion for selecting the backbone. To do this, each parent $p\in B$ gets associated with a (possibly empty) set of children~$\partial_p \subseteq G\setminus B$, each of which has at least one node in common with $p$ and is transmitted by exploiting its overlap with $p$. This is the fundamental mechanism by which the backbone~$B$ encodes structural information about the hypergraph~$G$, apart from being a subset of its hyperedges. A child~$c$ can be transmitted from the parent~$p$ by specifying: (i)~the size~$\abs{c}$ of the hyperedge, requiring $\log L$ bits; (ii)~the (non-zero) number of overlapping nodes~$\abs{p \cap c}$ with the parent~$p$, requiring $\log \min(\abs{p},\abs{c})$ bits; (iii)~the $\abs{p \cap c}$ nodes that overlap with $p$, requiring $\log {\abs{p}\choose \abs{p \cap c}}$ bits; (iv)~the $\abs{c}-\abs{p \cap c}$ nodes that do not overlap with $p$, requiring $\log {N-\abs{p}\choose \abs{c}-\abs{p \cap c}}$ bits. Putting this all together, the information content to transmit a child~$c \in G\setminus B$ from a parent~$p \in B$ is
\begin{align}\label{eq:Cond-ent}
H(c\vert p) &= \log L + \log \min (\abs{p},\abs{c}) \\
&~~~~~+ \log {\abs{p}\choose \abs{p \cap c}}{N-\abs{p}\choose \abs{c}-\abs{p \cap c}} \nonumber    
\end{align}
bits, and the information required to transmit the rest of $G$ given its backbone~$B$ is
\begin{align}
\mathcal{L}(G\vert B) = \sum_{p\in B}\sum_{c\in \partial_p}H(c\vert p).    
\end{align}
Analogous to the combinatorial entropy of Eq.~\eqref{eq:Hent}, Eq.~\eqref{eq:Cond-ent} provides a combinatorial conditional entropy for the child hyperedge~$c$ given its parent hyperedge~$p$.

The total description length to transmit an observed hypergraph~$G$ while using the backbone~$B\subseteq G$ for compression is then
\begin{align}\label{eq:Ltotalorig}
\mathcal{L}(G, B) &= \mathcal{L}(B) + \mathcal{L}(G\vert B)\\
&=\sum_{p\in B}\left[H(p)+\sum_{c\in \partial_p}H(c\vert p)\right].
\end{align}
Under the MDL principle, the best backbone~$B^\ast$ for summarizing the structure of the hypergraph~$G$ is the one that permits its greatest compression, or
\begin{align}\label{eq:argmin}
B^\ast = \argmin_{B\subseteq G}\left\{\mathcal{L}(G, B)\right\}.    
\end{align}
Minimizing $\mathcal{L}(G, B)$ thus provides a principled information-theoretic objective for identifying the structural backbone~$B^\ast$ that best captures the key structural regularities of $G$. If $B$ is too small or does not capture the important regularities within $G$, then the cost~$\mathcal{L}(G\vert B)$ will be high since the graph~$G$ does not share substantial hyperedge overlap with $B$. On the other hand, if $B$ is too large, then the cost~$\mathcal{L}(B)$ will be comparable to that of $G$ alone and the savings we get in the conditional entropy~$\mathcal{L}(G\vert B)$ cannot compensate for the high upfront cost to transmit $B$.

To determine the level of compression we can achieve for $G$ using the MDL-optimal backbone~$B^\ast$, we can compare the minimum description length value~$\mathcal{L}(G, B^\ast)$ with the description length~$\mathcal{L}(G, G)$ required to transmit $G$ without the aid of any backbone (i.e., $G$ is the backbone). The ratio of these two quantities gives a compression ratio,
\begin{align}\label{eq:comp-ratio}
\eta = \frac{\mathcal{L}(G, B^\ast)}{\mathcal{L}(G, G)},    
\end{align}
such that $\eta\in [0,1]$, with $\eta=1$ indicating that no compression is achievable using the backbone encoding and $\eta\to 0$ indicating substantial compression using the backbone.

In Fig.~\ref{fig:fig1_diagram} we show an example hypergraph with hyperedge orders $\ell\in\{2,3,4,5\}$ present, next to its inferred backbone obtained through solving Eq.~\eqref{eq:argmin}. The hypergraph backbone retains both global structural integrity and local connectivity without sacrificing compression ($\eta = 0.79$ in this example, using Eq.~\eqref{eq:comp-ratio}). 

\begin{figure}[t!]
\includegraphics[width=0.5\textwidth]{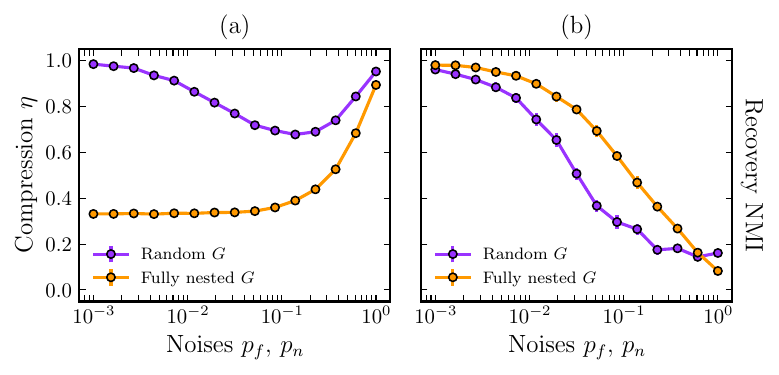}
\caption{
    \textbf{Backboning synthetic hypergraphs with tunable structure.}
    (a)~Compression ratio $\eta$ (Eq.~\eqref{eq:comp-ratio})
    against structural noise parameters $p_f$ and $p_n$ for the randomized hypergraph experiments described in
    Sec.~\ref{sec:unweighted}. 
    (b)~Recovery NMI scores comparing the inferred and planted ground truth backbones in these randomized hypergraph experiments.
}
\label{fig:fig2}
\end{figure}

It will be useful to rewrite the description length objective in Eq.~\eqref{eq:Ltotalorig} using a generalization of the reduced mutual information~(RMI)~\cite{newman2020improved,jerdee2024mutual} to hyperedge tuples $c$ and $p$. Using Eq.~\eqref{eq:Hent} and Eq.~\eqref{eq:Cond-ent}, the RMI between the parent hyperedge~$p$ and child hyperedge~$c$ is given by
\begin{align}
R(c,p) &= H(c) - H(c\vert p)\\
&= H(p) - H(p\vert c) \\
&= \log \frac{{N\choose \abs{p}}{N\choose \abs{c}}}{{N\choose \abs{p\cap c},\,\abs{p\setminus c},\,\abs{c\setminus p},\,N-\abs{p\cup c}}} - \log \text{min}(\abs{p},\abs{c}) \label{eq:rmi},
\end{align}
where $\abs{a\setminus b}=\abs{a}-\abs{a\cap b}$ is the size of the set difference between tuples $a$ and $b$, and $\abs{a\cup b}=\abs{a\setminus b}+\abs{b\setminus a}+\abs{a\cap b}$ is the size of their union. We can notice that Eq.~\eqref{eq:rmi} takes the form of a mutual information using the contingency table with entries~$\{\abs{p\cap c},\,\abs{p\setminus c},\,\abs{c\setminus p},\,N-\abs{p\cup c}\}$, and subtracting the transmission cost~$\log \text{min}(\abs{p},\abs{c})$ of this table gives the RMI. Equation~\eqref{eq:Ltotalorig} can then be written as
\begin{align}\label{eq:Ltotal-rmi}
\mathcal{L}(G, B) = \mathcal{L}(G,G) - \sum_{c\in G\setminus B}R(c,p(c)),    
\end{align}
where $p(c)$ is the parent of child hyperedge~$c$. We can interpret Eq.~\eqref{eq:Ltotal-rmi} as the information required to transmit $G$ by itself, minus the information shared by $G$ and the backbone~$B$.

Equation~\eqref{eq:Ltotal-rmi} tells us that the description length favors parent-child hyperedge pairs~$p,c$ with higher RMI values~$R(c,p)$. For fixed hyperedge sizes~$\abs{p},\abs{c}$, we can show that under mild conditions on the hyperedge sizes the RMI is monotonically increasing in the overlap~$\abs{p \cap c}$ and is always positive for $c\subset p$ (see Appendix~\ref{app:RMI-mono}). Thus, the description length objective encourages backbones that create parent-child pairs with higher overlap and nestedness, or in other words backbones that remove such structural redundancies from the original hypergraph. These are topological signatures unique to higher-order networks and absent in pairwise networks.

In a set of experiments on randomized synthetic hypergraphs, we consider two distinct starting configurations for applying structural noise. The first is a fully nested hypergraph on $10^4$ nodes consisting of $100$ 5-simplices with top faces whose nodes are drawn uniformly at random. The second is a random hypergraph on $10^3$ nodes consisting of $10$ parent hyperedges whose sizes are drawn uniformly at random, and for each parent we assign $10$ child hyperedges that start as identical copies of the parent. The level of structural redundancy is tuned for the fully nested hypergraph by replacing a fraction~$p_f$ of its hyperedges with hyperedges of the same sizes drawn uniformly at random (without replacement) from the set of $10^4$ nodes. For the random initial hypergraphs, noise is introduced by varying the probability~$p_n$ that each node of each child hyperedge---which start as an identical copy of their parent---is removed and replaced by a node drawn uniformly at random. All children that remain as copies of their parent after this process are removed, so that the input hypergraph~$G$ to the method does not have multi-edges. Results for the fully nested and random hypergraphs are averaged, respectively, over $10$ and $100$ realizations with error bars indicating $3$ standard errors in the mean.

In Fig.~\ref{fig:fig2}(a) we plot the compression ratio~$\eta$ (Eq.~\eqref{eq:comp-ratio}) for both experiments. In the fully nested hypergraph, the compression monotonically increases as a function of the noise level, starting below $\eta=0.5$ and approaching $\eta = 1$ as all the planted redundancy is removed through randomization. In Appendix~\ref{app:analytical-simp} we provide additional analytical context for these results, discussing the scaling of $\eta$ for zero noise in the fully nested hypergraph. For the experiments on randomized initial hypergraphs, we can see non-monotonicity in the compression~$\eta$ as a function of the noise. For low levels of noise, we observe $\eta\approx 1$, while at noise levels of $10^{-1}$ we see that $\eta$ bottoms out and begins to increase, reaching $\eta\approx 1$ again for high noise levels. The second half of this trend is consistent with that observed in the fully nested case---more noise makes compression based on structural overlaps more difficult, so the compression ratio approaches $1$. However, for low levels of noise we also see poor compression in this experiment. This is because each child hyperedge~$c$ does not alter any of its constituent nodes with probability~$(1-p_n)^{\abs{c}}$. Thus, for low values of $p_n$, many children end up being exact copies of their parent, so are discarded from the initial hypergraph. The remaining hypergraph in such cases becomes largely composed of the initial parent hyperedges, and thus there is not much redundancy to compress and $\eta\approx 1$ despite strong recovery accuracy.

In Fig.~\ref{fig:fig2}(b) we plot the hypergraph NMI \cite{felippe2026information} between the inferred backbone and the planted backbone (e.g., the top faces of the simplices or the randomized initial hyperedges, for the two test configurations). For both experiments, we observe a smooth decline in recovery accuracy as the structural noise parameters are increased, with the fully nested hypergraphs being easier to recover at most noise values.  

\begin{figure}[t!]
\includegraphics[width=0.5\textwidth]{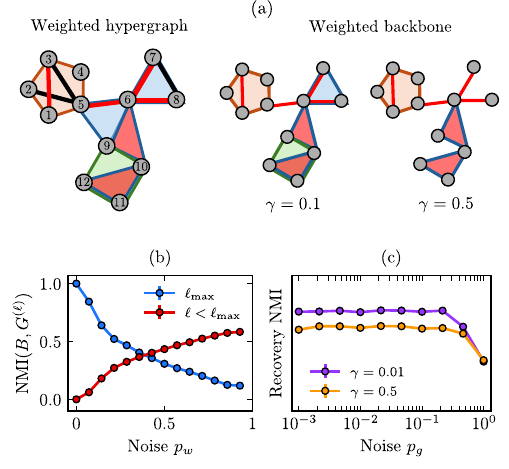}
\caption{
    \textbf{Backboning weighted hypergraphs.}
    (a)~Example weighted hypergraph consisting of hyperedges of orders $\ell\in \{2,3,4,5\}$. A subset of the hyperedges (highlighted in red) are assigned weight $w(e)=5$, while the remaining hyperedges are assigned weight $1$. Weighted backbones are inferred for this hypergraph by minimizing Eq.~\eqref{eq:DLweighted} at two different values of the parameter $\gamma$ (Eq.~\eqref{eq:gamma}), which tunes the relative impact of weight in determining hyperedge importance. (b)~Similarity between the inferred backbone and the top faces (blue) and other faces (red) of the fully nested hypergraph as their weight distributions are perturbed as described in Sec.~\ref{sec:weighted}. (c)~Recovery NMI for the randomized non-nested hypergraph experiments described in Sec.~\ref{sec:weighted}, for two different values of $\gamma$.
}
\label{fig:fig3}
\end{figure}

\section{Backboning weighted hypergraphs}
\label{sec:weighted}

The primary focus of existing pairwise network backboning techniques has been on edge weights~\cite{serrano2009extracting}. However, to our knowledge no hypergraph filtering method to date considers weights; existing approaches are based exclusively on topological redundancies such as nested hyperedges. Our backboning formulation naturally accommodates weighted hypergraphs by incorporating a prior over edge weights in the generative model for $G$, distinguishing between parent edges in the backbone~$B$ and child edges not in the backbone when assigning weights~\cite{kirkley2025fast}. Thus, our method operates on both topological and weight dimensions.

Let $w(e)$ be an edge weight assigned to the hyperedge~$e$, indicating for example the frequency or strength of interaction among the nodes in $e$. For simplicity, we will assume that $w(e)\geq 1$ is a positive integer, but the empirical Bayesian model we will describe can be used for both continuous and/or negative values of $w(e)$ as well, by utilizing a weight prior with the correct support. (Continuous weights additionally require the specification of a resolution parameter~\cite{hansen2001model}.)

According to the Kraft inequality, after we transmit the topology of the hypergraph $G$, we can transmit the weights~$w$ using a codelength of $-\log P(w)$ for any properly normalized distribution~$P(w)$. In this case, however, to allow for arbitrarily large weights, we no longer can use fixed-length codes corresponding to uniform distributions, since the support of $w$ is infinite. We thus have to choose a non-uniform weight distribution.  

To encode weights, we assume that the weight of each hyperedge is drawn independently from a distribution that depends only on whether the hyperedge belongs to the backbone. We denote this role by $b_e \in \{0,1\}$, with $b_e=1$ for parent (backbone) hyperedges and $b_e=0$ for child (non-backbone) hyperedges~\cite{kirkley2025fast}. The total description length then becomes
\begin{align}\label{eq:DLweighted}
\mathcal{L}_w(G,B) = \mathcal{L}(G,B) + \sum_{e \in G} \mathcal{L}(w(e), b_e),
\end{align}
where $\mathcal{L}(G,B)$ is the unweighted topological description length from Eq.~\eqref{eq:Ltotalorig} and $\mathcal{L}(w(e), b_e) = -\log P(w(e)\,\vert\, \theta_{b_e})$ is the codelength required to transmit the weight of $e$ given its role $b_e$, which fixes the parameters $\theta_{b_e}$ of the weight prior.

Let $\bar{w} = \frac{1}{|G|}\sum_{e \in G} w(e)$ be the empirical mean edge weight. Under an empirical Bayes constraint, we can impose that the average weight $\bar{w}$ is reproduced when averaging over both parent and child edges. Let $\mu_1 = \mathbb{E}[w \vert b=1]$ and $\mu_0 = \mathbb{E}[w \vert b=0]$ be the expected weights for parents and children, respectively. Define the ratio hyperparameter
\begin{align}\label{eq:gamma}
\gamma = \frac{\mu_0 - 1}{\mu_1 - 1} \in (0,1],
\end{align} 
where we subtract $1$ because $w=1$ is the minimum possible weight for the current case. (We can define $\gamma$ differently for continuous or negative weight distributions for specifying priors with the correct support.) When $\gamma = 1$, parents and children have the same expected weight. When $\gamma \to 0$, children have expected weight approaching $1$ (the minimum). Letting the weight priors be fixed before knowing which edges are in the backbone, we can enforce the global mean constraint~$\bar{w} - 1 = \frac{(\mu_1-1) + (\mu_0-1)}{2}$, giving
\begin{align}\label{eq:mu}
\mu_{b_e}(\gamma,\bar{w}) = 1 + \frac{2\gamma^{1-b_e}(\bar{w}-1)}{1+\gamma}
\end{align}
for the expected weight of each hyperedge role $b_e\in \{0,1\}$. Note that $\mu_{0}(\gamma,1)=\mu_{1}(\gamma,1)=1$ when the hypergraph is unweighted, and that $\mu_{0}(1,\bar{w})=\mu_{1}(1,\bar{w})=\bar{w}$ for $\gamma=1$. In the opposite limit, we have $\mu_{b_e}(\gamma,\bar{w})=1+2(\bar{w}-1)b_e$ for $\gamma\to 0^+$.

Here we consider both Poisson and geometric distributions for the weight prior~$P(w\,\vert\,\theta_{b_e})$ in our encoding. These families are chosen because they are fully specified by their means~$\theta_{b_e}=\mu_{b_e}$, allowing us to tune the relative importance of weights versus topology with a single parameter. However, weight prior families with more parameters~$\theta_{b_e}$ can also be utilized by matching higher moments of the empirical weight distribution with the moments of the weight prior. For the Poisson and geometric weight priors, $w(e)-1$ follows the respective distribution on $\{0,1,2,\dots\}$ with mean $\mu_{b_e}(\gamma,\bar{w})-1$. This results in the edge weight costs
\begin{align}
\mathcal{L}^{(\text{Pois})}(w(e),b_e) &= \frac{\mu_{b_e}-1}{\ln(2)} -(w(e)-1)\log(\mu_{b_e}-1) \nonumber\\
&~~~~~~~~~~+\log(w(e)-1)!, \\
\mathcal{L}^{(\text{Geom})}(w(e),b_e) &= \log (\mu_{b_e}) + (w(e)-1)\log\frac{\mu_{b_e}}{\mu_{b_e}-1}.
\end{align}
Several natural limits hold for both priors:
\begin{itemize}
    \item When $w(e)=1$ for all $e$, $\mathcal{L}(w(e),b_e)=0$ and Eq.~\eqref{eq:DLweighted} reduces to the unweighted objective in Eq.~\eqref{eq:Ltotalorig}.
    \item When $\gamma = 1$, $\mathcal{L}(w(e),0) = \mathcal{L}(w(e),1)$, so parents and children incur the same weight penalty in the description length of Eq.~\eqref{eq:DLweighted}, giving the unweighted objective in Eq.~\eqref{eq:Ltotalorig} with an irrelevant additive constant. 
    \item When $\gamma \to 0$ and $w(e)>1$, $\mathcal{L}(w(e),0) \to +\infty$ while $\mathcal{L}(w(e),1)$ stays finite, making it infinitely costly to assign a hyperedge with nontrivial weight as a child.
\end{itemize}

Thus, $\gamma\in (0,1]$ provides a principled tunable parameter for choosing how much to emphasize weight relative to topology when deciding the backbone using the weighted description length objective in Eq.~\eqref{eq:DLweighted}. As such, $\gamma=1$ reduces to the unweighted objective (plus an irrelevant constant), while $\gamma\approx 0$ places a harsh penalty on assigning hyperedges with nontrivial weight as children. 

The total weighted description length can be rearranged into a form analogous to Eq.~\eqref{eq:Ltotal-rmi}:
\begin{align}
\mathcal{L}_w(G,B) = \mathcal{L}_w(G,G) - \sum_{c \in G \setminus B} R_w(c, p(c)),
\end{align}
where
\begin{align}\label{eq:rmi-weighted}
R_w(c, p(c)) = R(c, p(c)) + \big [\mathcal{L}(w(c),1) - \mathcal{L}(w(c),0)\big ].
\end{align}

The term $\mathcal{L}(w(e),0) - \mathcal{L}(w(e),1)$ thus acts as a weight-dependent reward for making a hyperedge~$e$ a parent when $\gamma < 1$. For the two priors, this reward is given by
\begin{align}
\text{Poisson:}&~~ (w(e)-1)\log \frac{\mu_1-1}{\mu_0-1}-\frac{\mu_1-\mu_0}{\ln(2)},\\
\text{Geometric:}&~~ (w(e)-1)\log \frac{1-\mu_1^{-1}}{1-\mu_0^{-1}}-\log \frac{\mu_1}{\mu_0}.
\end{align}
Thus, the weighted objective provides a reward for adding an edge~$e$ to the backbone that is linear in the weight~$w(e)$ for both the Poisson and Geometric priors. 

Moreover, by setting $\mathcal{L}(w(e),1) = \mathcal{L}(w(e),0)$, we can identify the weight threshold $w^\ast$ below which the prior no longer encourages adding the weight to the backbone. For $\gamma <1$ we have
\begin{align}
\text{Poisson:}~~ w^\ast&=1+\frac{\frac{\mu_1-\mu_0}{\ln(2)}}{\log \frac{\mu_1-1}{\mu_0-1}}\\
&=1+\frac{2(\bar{w}-1)(1-\gamma)}{\ln(2)(1+\gamma)\log(1/\gamma)},\\
\text{Geometric:}~~ w^\ast&=1+\frac{\log \frac{\mu_1}{\mu_0}}{\log \frac{1-\mu_1^{-1}}{1-\mu_0^{-1}}}\\
&=1+\frac{\log\frac{\gamma+(2\bar{w}-1)}{1+(2\bar{w}-1)\gamma}}{\log (1/\gamma)-\log\frac{\gamma+(2\bar{w}-1)}{1+(2\bar{w}-1)\gamma}}.
\end{align}
One can show that $w^\ast(\gamma)$ is monotonically increasing in $\gamma\in (0,1)$ for both priors (see Appendix~\ref{app:w-mono}). Thus, in both cases there exists a threshold weight~$w^\ast(\gamma)$ above which edges are encouraged to become parents, and this threshold increases as $\gamma \to 1$ from below, indicating that the parameter $\gamma$ provides a more and more stringent criterion for adding edges to the backbone purely due to weight as $\gamma\to 1$. This further suggests that the weight prior provides a principled way to incorporate interaction frequencies into backbone extraction, with the single hyperparameter~$\gamma$ controlling the relative importance of weights versus topology. Technically, due to the addition of $\gamma$, the weighted hypergraph backboning objective is parametric. However, this is necessary for differentiating the backbone and non-backbone edges through weight priors unless one chooses different model families, which is a more invasive choice by the modeler than the choice of $\gamma$. 

Fig.~\ref{fig:fig3}(a) shows the backbones inferred for an example hypergraph (left) in which edges are endowed with one of two weights, $w(e)=5$ (highlighted in red) or $w(e)=1$, at two different values of $\gamma$ under the 
Poisson weight prior. We can see that for low $\gamma=0.1$ (middle), the backbone contains multiple edges with low weight, while for higher $\gamma=0.5$ (right), the majority of the backbone edges are those of high weight. This highlights the fact that larger values of $\gamma$ impose a more stringent weight threshold for keeping edges in the backbone without providing sufficient additional topological overlap beyond the high weight edges to justify their inclusion.        

To systematically examine the tradeoff among topology and edge weights in determining the structural backbones of hypergraphs, we apply our weighted backboning formulation to synthetic hypergraphs with controlled weight distributions on the hyperedges. We first generate fully nested hypergraphs on $100$ nodes with $100$ top faces whose size is $\ell_{\text{max}} = 6$. We then draw the weight of each hyperedge~$e$ according to a geometric distribution with mean~$1+(1-p_w)\mu$ if $e\in B$ and a geometric distribution with mean~$1+p_w\mu$ otherwise, for $\mu=0.5$ and $p_w\in [0,1]$ a noise parameter. Increasing the parameter $p_w$ thus allows us to increase the discrepancy between the weight distribution on the top simplex faces and on the nested faces, increasing the signal present for including the nested faces in the backbone. To impose mis-specification on the weight model for assessing the robustness of this choice in our algorithm, we infer the backbones using the Poisson weight prior (with $\gamma=0.5$), despite the geometric distributions on the weights. All results were averaged over $10$ trials with error bars representing $2$ standard errors in the mean of the trials.

In Fig.~\ref{fig:fig3}(b) we plot the recovery NMI as a function of the noise~$p_w$ in this experiment. We can see that for $p_w=0$, the inferred backbone is identical to the top faces of the simplex, since these faces are both topologically significant and have systematically higher weights than the nested faces. As $p_w$ increases, however, the inferred backbones become increasingly dominated by nested hyperedges, since there are many more of these hyperedges and their weights are systematically higher than those of the top faces for $p_w>0.5$. For $p_w=1$, we observe a reversal in the composition of the backbone, with most hyperedges being those that are nested within the top faces. This is because these nested hyperedges are dominant according to weight in this regime. 

We then run a second set of experiments to identify the impact of the choice of $\gamma$ on recovery of planted backbone structure in weighted systems. In these experiments, we start with a random hypergraph on $100$ nodes generated by the procedure described in Sec.~\ref{sec:unweighted}, setting the number of parents to be $20$ and the number of children for each parent to be $2$. We then set the topological noise level to be $p_n=0.5$, so that there was only a weak topological signal present for recovery, and set $w(e)=1$ for all child hyperedges. For parent hyperedges, we draw $w(e)$ from a geometric distribution with mean $1/p_g$, where $p_g$ tunes the level of weight noise in the system. Lower values of $p_g$ create stronger discrepancies between the parents and the children, making the parents easier to recover as the backbone, while $p_g=1$ forces the parents and children to have the same weights so that only the (weak) topological overlap can be used to determine the backbone. In Fig.~\ref{fig:fig3}(c) we show the results of these experiments, computing the similarity between the inferred backbones and the planted parent hyperedges as a function of $p_g$. We find that recovery accuracy remains consistently high for low $p_g$ but begins to decline for $p_g\approx 0.2$. The NMI reaches its lowest level for $p_g=1$, when backbone recovery is driven solely by the noisy topological signal in the hypergraph.

Similar to network community detection, minimizing $\mathcal{L}(G,B)$ or its weighted counterpart~$\mathcal{L}_w(G,B)$ over all subsets~$B \subseteq G$ is an intrinsically complex combinatorial optimization problem. In Appendix~\ref{sec:optimization} we describe a few efficient approaches for approximating the solution to this problem, in the same spirit as greedy approaches for other network inference tasks~\cite{clauset2004finding,blondel2008fast,kirkley2024identifying}. Both approaches apply to the unweighted objective~$\mathcal{L}(G,B)$ in Eq.~\eqref{eq:Ltotalorig} and the weighted objective~$\mathcal{L}_w(G,B)$ in Eq.~\eqref{eq:DLweighted}. We verify the theoretical runtime complexity estimates of our method in this appendix as well.

In Appendix~\ref{app:local-derivation} we describe a variant of our backboning method that focuses on each node neighborhood (i.e., set of hyperedges including a node) separately. This allows for the advantages of the disparity filter~\cite{serrano2009extracting} and its variants to be realized for hypergraphs with heterogeneous node connectivity. We repeat the experiments above for this alternative backboning approach in the same appendix.

%%%%%%%%%%%%%%%%%%%%%%%%%%%%%%%%%%%%%%%%%%%%%%%%%%%
%%%%%%%%%%%%%%%%%Results%%%%%%%%%%%%%%%%%%%%%%%%%%%
%%%%%%%%%%%%%%%%%%%%%%%%%%%%%%%%%%%%%%%%%%%%%%%%%%%

\section{Backboning real-world hypergraphs}
\label{sec:empirical}

We apply the proposed backboning method to a range of empirical higher-order datasets, obtained from the Hypergraphx-data repository \cite{lotito2026hypergraphx}. These datasets come from a variety of real-world processes including scientific coauthorship, email exchanges, human contacts, and drug documentation among others.

We apply the unweighted higher-order backboning method of Sec.~\ref{sec:unweighted} to all of these empirical hypergraphs and report the sizes of the original hypergraph $\abs{G}$, backbone $\abs{B}$, and compression ratio $\eta$ (Eq.~\eqref{eq:comp-ratio}) in Table~\ref{tab:datasets}. For each system we run both backbone optimization schemes (``edge''- and ``node''-based, following Appendix~\ref{sec:optimization}), choosing the one that provides the lower description length for each network. We find that the edge-based scheme provides better compression in the majority of cases. 

\begin{table}[hb!]
\centering
\tiny
\rowcolors{2}{white}{gray!10}
\resizebox{0.5\textwidth}{!}{
\begin{tabular}{lcccc}
    \hline\toprule\toprule
    \textbf{Dataset} & \multicolumn{1}{c}{$N$} & \multicolumn{1}{c}{$|G|$} & \multicolumn{1}{c}{$|B|$} & \multicolumn{1}{c}{$\eta$} \\
    \toprule\toprule\hline
    coauth-mag-geology\_1980 & 1674 & 903 & 310 & 0.80 \\
    coauth-mag-geology\_1981 & 1075 & 547 & 192 & 0.82 \\
    coauth-mag-geology\_1982 & 1878 & 987 & 345 & 0.80 \\
    coauth-mag-geology\_1983 & 1734 & 883 & 313 & 0.81 \\
    contact-high-school & 327 & 7818 & 2223 & 0.65 \\
    contact-primary-school & 242 & 12704 & 3579 & 0.70 \\
    dawn & 2290 & 138742 & 52704 & 0.83 \\
    email-enron & 143 & 1459 & 309 & 0.61 \\
    email-eu & 986 & 24520 & 5484 & 0.60 \\
    hospital-lyon & 75 & 1824 & 545 & 0.69 \\
    hypertext-conference & 113 & 2434 & 691 & 0.74 \\
    invs13 & 92 & 787 & 222 & 0.77 \\
    invs15 & 217 & 4909 & 1415 & 0.70 \\
    kaggle-whats-cooking & 6714 & 39224 & 8918 & 0.78 \\
    malawi-village & 84 & 431 & 125 & 0.68 \\
    ndc-classes & 628 & 796 & 226 & 0.54 \\
    ndc-substances & 3414 & 6471 & 1549 & 0.60 \\
    science-gallery & 410 & 3350 & 924 & 0.64 \\
    sfhh-conference & 403 & 10541 & 2989 & 0.70 \\
    tags-ask-ubuntu & 3021 & 145053 & 43770 & 0.73 \\
    tags-math-sx & 1627 & 169259 & 57071 & 0.75 \\
    \bottomrule
\hline
\end{tabular}
}
    \caption{Higher-order backboning of real-world hypergraphs.
    }
    \label{tab:datasets}
\end{table}

\begin{figure}[ht!]
\includegraphics[width=0.50\textwidth]{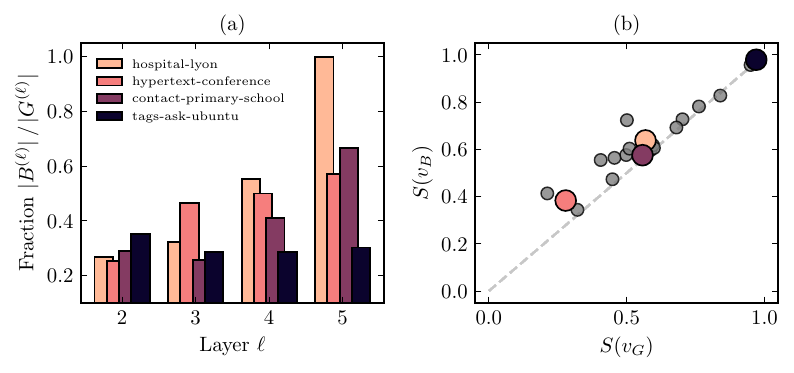}
\vspace{-2em}

\caption{
    \textbf{Layerwise heterogeneity in empirical backbones.}
    (a)~Fraction of edges kept in each layer $\ell$ of the backbone, for a small set of empirical network examples. We can see substantial heterogeneity, in contrast with the binary layer removal enforced by structural reducibility \cite{kirkley2025structural}.
    (b)~Normalized entropy of edge sizes (Eq.~\eqref{eq:layer-ent}) for the original hypergraph and its backbone, highlighting the four selected real-world hypergraphs of panel (a).
}
\label{fig:empirical-heterogeneity}
\end{figure}

Table~\ref{tab:datasets} shows that all empirical hypergraphs contain structural redundancies that can be effectively removed through the backboning method. We consistently find that the backbone $B$ contains roughly one-quarter to one-third of the hyperedges present in the original graph $G$, providing meaningful sparsification of the network. The compression ratios $\eta$ range from $\eta=0.54$ to $\eta=0.83$, indicating that substantial compression of each hypergraph is possible by exploiting overlaps and nestedness to produce a more compact representation of the system's higher-order structure.   
In Fig.~\ref{fig:empirical-heterogeneity} we show the heterogeneity in hyperedge sizes produced by the backboning procedure across the empirical hypergraph corpus, contrasting with the binary layer removal associated with structural reducibility \cite{kirkley2025structural}. In Fig.~\ref{fig:empirical-heterogeneity}(a) we show the fraction of hyperedges retained in each layer $\ell$ for a selected set of empirical networks, illustrating the diversity of pruning behaviors possible under the proposed backboning method.

In Fig.~\ref{fig:empirical-heterogeneity}(b), we plot the normalized Shannon entropy of the layer sizes present for each hypergraph, measured as
\begin{align}\label{eq:layer-ent}
S(v_G) = -\frac{1}{\log L}\sum_{\ell}v_G(\ell) \log v_G(\ell),    
\end{align}
with 
\begin{align}
v_G=\left\{\abs{G^{(2)}},\abs{G^{(3)}},\dots,\abs{G^{(\ell_{\text{max}})}}\right\}/\abs{G}    
\end{align}
the fractions of hyperedges contained in each layer $G^{(\ell)}$ and $L$ the number of layers in $G$. The fractional layer sizes $v_B$ are defined analogously.  Figure~\ref{fig:empirical-heterogeneity}(b) shows that higher-order backboning consistently produces more balanced layer sizes, as the normalized layer size entropy of Eq.~\eqref{eq:layer-ent} increases in nearly all systems studied. This homogenization of hyperedge sizes is because empirical hypergraphs typically have more lower-order interactions than higher-order interactions, and higher-order hyperedges are more likely to be retained when nestedness and overlap are used for compression. (Larger hyperedges tend to overlap with more hyperedges than smaller hyperedges.) Thus, the larger layers of lower-order hyperedges tend to be more aggressivley pruned under higher-order backboning. 

We also apply the local backboning method of Appendix~\ref{app:local-derivation} to these datasets, presented in Table~\ref{tab:local} in the same appendix. Meanwhile, in Appendix~\ref{app:weighted-real} we apply the weighted backboning method of Sec.~\ref{sec:weighted} to the subset of these empirical hypergraphs that have temporal information, assigning each hyperedge weight to be its number of occurrences over time. Finally, in Appendix~\ref{app:reducibility}, we compare hypergraph backboning with the related idea of hypergraph order reducibility \cite{lucas2026reducibility,kirkley2025structural}, providing an overview of their conceptual and practical distinctions.

\section{Conclusion}
In this paper, we introduce a principled information-theoretic framework based on the minimum description length (MDL) principle for extracting structural backbones from hypergraphs. Our method identifies minimal representations of higher-order interactions by compressing nested and redundant hyperedges while preserving essential structural information. This framework is fully nonparametric for unweighted hypergraphs, and naturally accommodates edge weights through an empirical Bayesian extension with a single tunable parameter balancing the tradeoff between topology and edge weight. Through validation on synthetic systems with controlled nestedness and redundancy, we demonstrate that our method consistently recovers planted structural regularities, and in applications to empirical hypergraphs from diverse domains we find substantial sparsification of real-world higher-order systems. Our method enables more efficient and interpretable analyses of higher-order interactions in complex systems, contributing to an emerging body of work developing principled information theoretic methods for understanding and inferring hypergraph structure from empirical relational data~\cite{young2021hypergraph,lizotte2023hypergraph,wegner2014subgraph,wegner2024nonparametric,kirkley2024inference,kirkley2025structural,felippe2026information}. 

There are several ways in which this work can be extended. First, optimization with simulated annealing or other more sophisticated Monte Carlo methods could allow for improved minimization of the description length objective, though at the cost of computational efficiency. Additionally, extensions to temporal hypergraphs would allow for greater flexibility in downstream applications to contagion and other asymmetric time-dependent dynamical processes through the identification of persistent and transient higher-order structures. 
Our work provides new insights into the organizational principles of higher-order networks, exploiting redundancies to construct parsimonious representations of hypergraphs that result in minimal loss of structural information.

\section*{Acknowledgments}

A.K.~acknowledges support from the HKU-100 Start Up Fund. F.B.~acknowledges support from the Austrian Science Fund (FWF) through projects 10.55776/PAT1052824 and 10.55776/PAT1652425.\\

\section*{Data and code availability}

The empirical datasets analyzed in this paper are openly available via Hypergraphx-data~\cite{lotito2026hypergraphx}, and the code is distributed within the Hypergraphx library~\cite{lotito2023hypergraphx}.

\clearpage
\appendix
\onecolumngrid

\section{Proof of RMI properties}
\label{app:RMI-mono}

The RMI in Eq.~\eqref{eq:rmi} can be equivalently written in terms of the overlap~$\abs{p\cap c}$ as
\begin{align}
R(c,p) = \log \frac{{N\choose \abs{p}}{N\choose \abs{c}}}{{N\choose \abs{p\cap c},\,\abs{p}-\abs{p\cap c},\,\abs{c}-\abs{p\cap c},\,N-\abs{p}-\abs{c}+\abs{p\cap c}}} - \log \min(\abs{p},\abs{c}).    
\end{align}
To show that the RMI is monotonically increasing in the overlap~$\abs{p\cap c}$ for fixed $\abs{p},\abs{c}$, we can equivalently show that the function
\begin{align}
f(x) = \log\big[x!(\abs{p}-x)!(\abs{c}-x)!(N-\abs{p}-\abs{c}+x)!\big]
\end{align}
is monotonically increasing in $x$ for $x\in \big[0,\min(\abs{p},\abs{c})\big]$. We have that
\begin{align}
f(x+1)-f(x) = \log \frac{(x+1)(N-\abs{p}-\abs{c}+x+1)}{(\abs{p}-x)(\abs{c}-x)},   
\end{align}
and so positive monotonicity is equivalent to the condition
\begin{align}
(x+1)(N-\abs{p}-\abs{c}+x+1) > (\abs{p}-x)(\abs{c}-x).    
\end{align}
Expanding through and simplifying, this is equivalent to
\begin{align}
(N+2)x+(N+1)>\abs{p}\abs{c}+\abs{p}+\abs{c}.    
\end{align}
Now, the lowest possible value for the LHS within the domain of interest is $x=0$, so the monotonicity condition will be satisfied for all overlap values~$x\in \big[0,\min(\abs{p},\abs{c})\big]$ so long as $N>\abs{p}\abs{c}+\abs{p}+\abs{c}-1$, which will be true for all hyperedges~$p,c$ that are small compared to the total number of nodes~$N$. Thus, the RMI is monotonically increasing in the hyperedge overlap, under a weak condition upper bounding the hyperedge sizes relative to the total number of nodes in the hypergraph.

\medskip

Note also that for $c\subset p$, we have 
\begin{align}
R(c,p)=\log \frac{{N\choose \abs{c}}{N\choose \abs{p}}}{\abs{c}{N\choose \abs{c},\,\abs{p}-\abs{c},\,N-\abs{p}}},
\end{align}
which is greater than zero for
\begin{align}
{N\choose \abs{c}}{N\choose \abs{p}} > \abs{c}{N\choose \abs{c},\abs{p}-\abs{c},N-\abs{p}}.    
\end{align}
This condition simplifies to
\begin{align}
{N\choose \abs{c}} > \abs{c} {\abs{p}\choose \abs{c}}.    
\end{align}
Now, for a similarly weak condition~$N>e\abs{p}\abs{c}^{1/\abs{c}}$ on the hyperedges sizes, we have
\begin{align}
\abs{c} {\abs{p}\choose \abs{c}} &\leq \abs{c}\left(\frac{e\abs{p}}{\abs{c}}\right)^{\abs{c}}
= \left(\frac{e\abs{p}\abs{c}^{1/\abs{c}}}{\abs{c}}\right)^{\abs{c}} < \left(\frac{N}{\abs{c}}\right)^{\abs{c}} \leq {N\choose \abs{c}},
\end{align}
where we have used the inequality~\cite{cover2012elements}
\begin{align}
\left( \frac{n}{k} \right)^k \leq {n \choose k} \leq \left( \frac{e n}{k} \right)^k.
\end{align}
The inequality required for positivity of $R(c,p)$ is thus satisfied under the condition~$N>e\abs{p}\abs{c}^{1/\abs{c}}$. Thus, the RMI is positive for all $c\subset p$ under the same condition, which is comparatively even milder than the condition for monotonicity.

\newpage

\section{Compression of a simplicial complex}
\label{app:analytical-simp} 

Consider a simplicial complex~$G$ consisting of $S$ disjoint simplices on $K\geq 3$ nodes, such that $N=KS$ is the total number of nodes in the hypergraph~$G$ and $L=K-1$ is the number of unique hyperedge orders. Let $p$ denote the top face of one of the simplices, with size $\abs{p}=K$, such that all other edges~$c$ in that simplex are subsets of $p$. For a moderate number of simplices~$S\geq K^{1+1/\abs{c}}$, we have
\begin{align}
H(c\vert p) &= \log L + \log K + \log {K\choose \abs{c}}\\
&\leq \log L + \log K + |c| \log\left( \frac{e K}{|c|} \right)\\
&\leq \log L + \log K + |c| \log\left( \frac{K^2}{|c|} \right) \quad \text{(since $K > e$)}\\
&= \log L + (2|c|+1)\log K - |c|\log|c|\\
&\leq \log L + |c|\log S + |c|\log K - |c|\log|c| \quad \text{(since $S\geq K^{1+1/\abs{c}}$)}\\
&= \log L + |c| \log\left( \frac{K S}{|c|} \right)\\
&= \log L + |c| \log\left( \frac{N}{|c|} \right)\\
&\leq \log L + \log {N\choose \abs{c}}\\
&= H(c),
\end{align}
where we have again used the inequality
\begin{align}
\left( \frac{n}{k} \right)^k \leq {n \choose k} \leq \left( \frac{e n}{k} \right)^k.
\end{align}
Thus, any hyperedges~$c$ that are not top faces will never be contained in an MDL-optimal backbone, since we can always reduce the total description length by making $c$ a child to its simplex's top face. Consequently, the MDL-optimal backbone of such a simplicial complex can only contain the top faces of size $K$. Since these hyperedges are disjoint and they cannot be parents to one another, the MDL-optimal backbone $B^\ast$ must simply be this set of $S$ top faces. 

\medskip

\begin{figure*}[h!]
\includegraphics[width=1.00\textwidth]{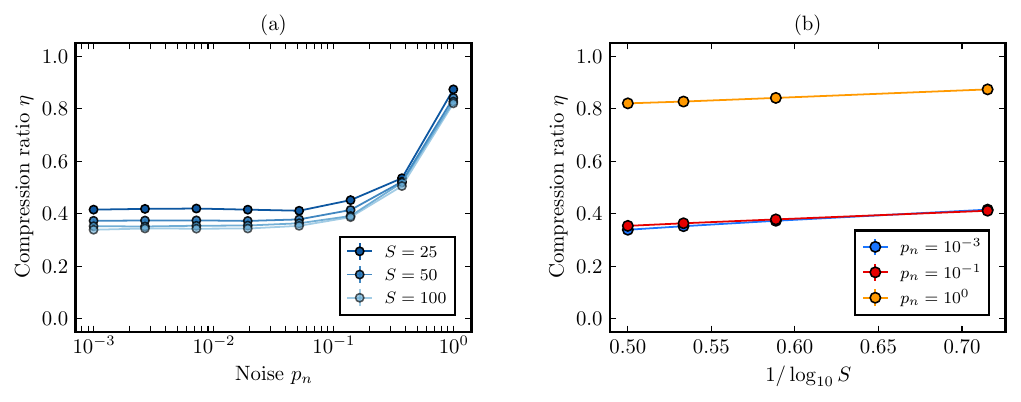}
\caption{
    \textbf{Compression of a simplicial complex.}
    (a)~Compression ratio $\eta$ (Eq.~\eqref{eq:comp-ratio}) against noise $p_n$ in the experiments of Sec.~\ref{sec:unweighted}, for simplicial complexes with a varying number of top faces $S$.
    (b)~$\eta$ versus $1/\log{S}$ for the same set of experiments, numerically illustrating the relation $\eta\sim 1/\log S$ at various noise levels $p_n$.
}
\label{fig:fig_1logS}
\end{figure*}

The minimum description length associated with each simplex with top face~$p$ and nested children~$c$ is then
\begin{align}
\ell(K) &= H(p) + \sum_{c\subset p}H(c\vert p)\\
&= \left[\log (K-1) + \log {KS \choose K}\right] + \sum_{k=2}^{K-1}{K\choose k}\left[\log (K-1) + \log K + \log {K\choose k}\right] \\
&= \log (K-1)+[2^K-K-2]\log [K(K-1)] + \log {KS \choose K} + \sum_{k=2}^{K-1}{K\choose k}\log {K\choose k}\\
&\sim O(K\cdot 2^K).
\end{align}
Meanwhile, the naive description length for the simplex obtained by placing all its edges in the backbone is
\begin{align}
\ell_0(K) = [2^K-K-1]\log (K-1) + \sum_{k=2}^{K}{K\choose k}\log {KS\choose k} \sim O(K\cdot 2^K\log S)     
\end{align}
Now, using Eq.~\eqref{eq:Ltotalorig}, the compression ratio for this simplicial complex scales as
\begin{align}
\eta &= \frac{S\times \ell(K)}{S\times \ell_0(K)} \sim \frac{1}{\log S},      
\end{align}
which vanishes for large $S$ but does not depend to leading order on the number of nodes~$K$ in each simplex. Figure~\ref{fig:fig_1logS} validates this result numerically by replicating the random hypergraph experiments of Sec.~\ref{sec:unweighted} for a simplicial complex with $1000$ nodes and varying number of top faces of maximum size $\ell=6$. Panel~(a) shows the full result for $S=25$, $50$, and $100$ at different noise perturbations $p_n$, that is, the probability that each node of each child hyperedge is removed and replaced by a node drawn uniformly at random. Panel~(b) shows the compression ratio $\eta$ versus $1/\log{S}$, exhibiting the linear scaling relation conjectured above.

\newpage
\section{Proof of $w^\ast(\gamma)$ monotonicity}
\label{app:w-mono}

As shown in Sec.~\ref{sec:weighted}, the threshold weight $w^\ast$ above which the weighted description length objective in Eq.~\eqref{eq:DLweighted} begins to reward adding an edge to the backbone $B$ is given by
\begin{align}
w_{\text{pois}}^\ast(\gamma) = 1+\frac{2(\bar{w}-1)(1-\gamma)}{\ln(2)(1+\gamma)\log(1/\gamma)}
\end{align}
for the Poisson weight prior, and
\begin{align}
 w_{\text{geom}}^\ast(\gamma) = 1+\frac{\log\frac{\gamma+(2\bar{w}-1)}{1+(2\bar{w}-1)\gamma}}{\log (1/\gamma)-\log\frac{\gamma+(2\bar{w}-1)}{1+(2\bar{w}-1)\gamma}}   
\end{align}

\medskip

for the geometric weight prior. We can show that both of these thresholds are monotonically increasing in $\gamma\in (0,1)$, indicating that as $\gamma$ increases, the weighted description length objective imposes a harsher and harsher criterion for including a hyperedge in the backbone purely based on its weight.

For the Poisson case, we have that setting $\pder{w_{\text{pois}}^\ast(\gamma)}{\gamma}>0$ gives the condition
\begin{align}
2 \log \gamma -\gamma +\gamma^{-1} > 0.
\end{align}
Letting $t=\gamma^{-1}\in (1,\infty)$ and $g(t)=f(\gamma)=f(1/t)$, we have
\begin{align}
g(t) &= t - t^{-1} - 2\log t\\
g'(t) &= 1 + t^{-2} - 2t^{-1}\\
&= t^{-2}(t-1)^2.
\end{align}
We have that $g'(t)>0$ for all $t>1$ and $g(1)=0$, thus $g(t)>0$ for all $t>1$. Since $g(t)=f(1/t)$, we therefore know that $f(1/t)>0$ for all $t>1$, or equivalently that $f(\gamma)>0$ for all $t\in (0,1)$. Thus, $w^\ast_{\text{pois}}(\gamma)$ is monotonically increasing for $\gamma\in (0,1)$.  

\medskip

For the geometric case, we have that setting $\pder{w_{\text{geom}}^\ast(\gamma)}{\gamma}>0$ gives the condition
\begin{align}
\gamma^{-1}\log \frac{\gamma+a}{1+a\gamma} > \frac{1-a^2}{(\gamma+a)(1+a\gamma)}\log\gamma,
\end{align}
where $a=2\bar{w}-1 > 1$ for weighted hypergraphs. Letting
\begin{align}
f(a) = \gamma^{-1}\log \frac{\gamma+a}{1+a\gamma} - \frac{1-a^2}{(\gamma+a)(1+a\gamma)}\log\gamma,    
\end{align}
we have that
\begin{align}
f'(a) = \frac{1-\gamma^2}{\gamma(\gamma+a)(1+a\gamma)}  + \frac{(1+a^2)(1+\gamma^2)+2a\gamma}{(\gamma+a)^2(1+a\gamma)^2}.  
\end{align}
For $\gamma\in (0,1)$ and $a>1$, we then have that $f'(a)>0$. We also have that since $f(1)=0$, then the inequality $f(a)>0$ must hold for all $a>1$. Thus, $w^\ast_{\text{geom}}(\gamma)$ is monotonically increasing for $\gamma\in (0,1)$. 

\newpage
\section{Optimization}
\label{sec:optimization}

In this Appendix we describe two efficient greedy approaches for approximating the MDL solution to the unweighted objective $\mathcal{L}(G,B)$ in Eq.~\eqref{eq:Ltotalorig} and the weighted objective $\mathcal{L}_w(G,B)$ in Eq.~\eqref{eq:DLweighted}. 

\medskip

Note that, as described in Sec.~\ref{sec:unweighted}, parent-child relationships $(c,p(c))$ must have at least a node overlap of at least $\abs{c\cap p(c)}=1$ according to the transmission objective. Thus, prior to optimization, it is useful to form the \emph{intersection graph} of $G$, defined as
\begin{align}
\text{Int}(G)=\{(e_1,e_2)\in G: e_1\neq e_2,~\abs{e_1\cap e_2} \geq 1\}.    
\end{align}

\medskip

$\text{Int}(G)$ is formed by associating each hyperedge~$e\in G$ with a ``node'' and connecting all pairs of adjacent ``nodes'' (i.e., pairs of hyperedges with at least one node in common). Any parent-child relationship~$(c,p(c))$ must then take place along an edge of $\text{Int}(G)$. This structure substantially reduces the number of potential pairs of parent-child we must consider in the objective. We know that the final parent-child relationships must form a partition of $\text{Int}(G)$ into disjoint stars, since each child~$c$ can only have one parent~$p(c)$ and cannot have any children, while each parent~$p$ cannot have a parent.

Given the intersection graph~$\text{Int}(G)$, we can approximately minimize the description length~$\mathcal{L}(G,B)$ over subsets~$B\subseteq G$ using a greedy ``node'' addition process where we start with an empty backbone~$B=\{\phi\}$ and iteratively pick the hyperedge~$e \in G\setminus B$ that produces the greatest decease to the description length, adding it to the backbone~$B$. A standard greedy procedure here would add all hyperedges until $B=G$ then pick the configuration with the lowest description length. However, this is computationally costly, as it requires $O(\abs{G \setminus B}^2)$ checks to find the best hyperedge to add at each step, since the change in description length from adding $e$ to the backbone~$B$ depends on which nodes are already in the backbone. A much faster procedure (thus the one we employ here) is to initially classify all nodes as being neither children nor parents, and add hyperedges to the backbone greedily until every hyperedge has been given a definitive role (i.e., is a parent or child in the current configuration). This process works for both the unweighted case (with the symmetric RMI of Eq.~\eqref{eq:rmi}) and the weighted case (with the asymmetric RMI of Eq.~\eqref{eq:rmi-weighted}). 

By viewing the RMI values as ``edge'' weights in the intersection graph~$\text{Int}(G)$, we can also perform optimization using a fast greedy approach that picks the best \emph{edge} in the intersection graph at each iteration rather than the best \emph{node}. At each iteration, we can add the ``edge'' (parent-child relationship) that most decreases the description length while preserving the required star partition structure. Moves that do not preserve this star structure are ignored, as they violate the condition that each child has a single parent and nodes cannot be simultaneously parents and children. 

Prior to running this greedy ``edge'' addition algorithm we need to first take all isolated hyperedges (ones with no overlaps) and put them as parents with no children. We also need to postprocess any remaining pairs of overlapping hyperedges in which neither hyperedge has overlaps with other hyperedges. For these cases, it does not matter for the description length which of the two hyperedges in the pair gets chosen for the backbone, and we can just choose the larger hyperedge of the two. (If they are the same size, we can pick one at random.) 

For a given hypergraph, we can then apply both the greedy ``node'' and greedy ``edge'' minimization approaches, picking the one that gives the lower description length. The computational bottleneck of both approaches is constructing the intersection graph~$\text{Int}(G)$. This procedure has a runtime complexity of $O(\abs{G}^2)$ if we check all pairs~$e_1,e_2 \in G\times G$ of hyperedges in $G$. However, we can also instead construct each node $i$'s neighborhood of adjacent hyperedges~$G_i=\{e\in G:i\in e\}$ with runtime complexity~$O\left(\sum_{e\in G}\abs{e}\right)$, then add all pairs within each node $i$'s neighborhood to the intersection graph, giving a runtime complexity of $O\left(\sum_{i=1}^{N}\abs{G_i}^2\right)$ for constructing $\text{Int}(G)$. This latter approach is faster in most practical cases of interest: In the worst case, for very dense hypergraphs, we can have a runtime complexity that scales as
$O\left(\sum_{i=1}^{N}\abs{G_i}^2\right) \sim O(N\abs{G}^2)$, but in the best case---if each node has a number of adjacent hyperedges~$k$ that is a constant independent of network size---the runtime complexity reduces to $O(Nk^2)\sim O(N)$. When either approach is computationally prohibitive (i.e., for large and dense hypergraphs), one can sample a set of random hyperedge pairs~$e_1,e_2\in G_i$ incident to node~$i$ uniformly at random, which reduces the runtime complexity to $O(Ns^2)$, where $s$ is the number of pairs to be sampled if ${\abs{G_i}\choose 2}>s$.

\begin{figure*}[h!]
\includegraphics[width=1.00\textwidth]{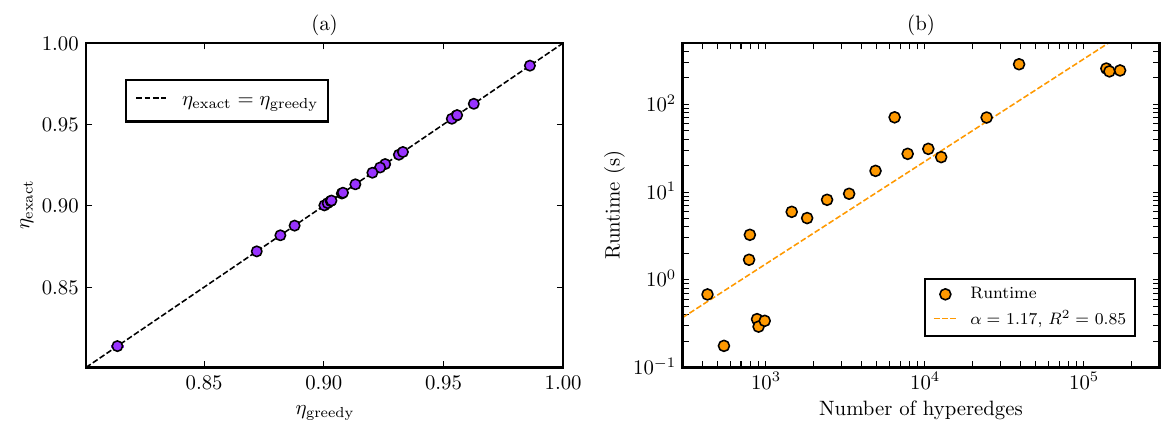}
\caption{
    \textbf{Optimization algorithm performance.}
    (a)~Compression ratio $\eta$ (Eq.~\eqref{eq:comp-ratio}) obtained from the greedy ``edge''-based algorithm and exact enumeration over all possible backbones, for small hypergraph samples extracted from empirical datasets via random walks over hyperedges. Each sample contained approximately $L=15$ hyperedges over $100$ realizations.
    (b)~Runtime versus hypergraph size $\abs{G}$ for the greedy method, for all empirical systems studied.
}
\label{fig:fig_runtime}
\end{figure*}

Fig.~\ref{fig:fig_runtime} shows numerical results demonstrating the strong performance of the greedy ``edge''-based algorithm, which we find has superior compression performance when compared to the ``node''-based approach, for nearly all empirical networks. In Fig.~\ref{fig:fig_runtime}(a) we show the compression ratio $\eta$ (Eq.~\eqref{eq:comp-ratio}) obtained from the greedy method along with the compression ratio obtained using exact enumeration over all possible backbones, for small hypergraph samples extracted from empirical datasets by collecting the hyperedges traversed via random walks over the hypergraph starting at random seed nodes. Each sample contained approximately $\abs{G}=15$ hyperedges over $100$ random walk realizations. We can see that for these cases in which we can directly evaluate the performance of the greedy method, it achieves compression that is indistinguishable from the optimal compression level under the description length objective. 

In Fig.~\ref{fig:fig_runtime}(b) we plot the runtime of the greedy algorithm when applied to all systems in the empirical hypergraph corpus. We also run an OLS regression $\log(\text{Runtime})\sim \alpha\log \abs{G}$, finding an exponent of $\alpha \approx 1.17$ with $R^2=0.85$. We observe that the runtime scales slightly superlinearly with the number of nodes $N$, indicating that in practice the algorithm is operating somewhere within the $O(Nk^2)$ and $O(N\abs{G}^2)$ runtime complexity estimates, while being closer to the former. The absolute runtimes are also quite modest for these empirical systems, despite the variety in hypergraph sizes, never exceeding $6$ minutes using a simple Python implementation.

\newpage
\section{Local backboning}
\label{app:local-derivation} 

Using the same framework as in Sec.~\ref{sec:unweighted} and Sec.~\ref{sec:weighted}, we can derive a backboning method that works locally at the node-level, similar to the disparity filter and its variants~\cite{serrano2009extracting,marcaccioli2019polya,kirkley2025fast}. 

Define each node ``neighborhood''~$G_i=\{e: e\in G,~i\in e\}$ as the set of all hyperedges in $G$ containing node~$i$. We can then construct a local backboning objective for each neighborhood~$G_i$, analogous to Eq.~\eqref{eq:argmin} of the main text,
\begin{align}
B_i^\ast = \argmin_{B_i\subseteq G_i}\left\{\mathcal{L}(G_i, B_i)\right\},    
\end{align}
where the subgraph~$G_i$ is considered in isolation. For weighted hypergraphs, we can just swap $\mathcal{L}_w$ from Eq.~\eqref{eq:DLweighted} for $\mathcal{L}$ in Eq.~\eqref{eq:Ltotalorig}.

We can then combine the local backbones~$\{B^\ast_i\}_{i=1}^{N}$ to create a local backbone $B^{(\text{local})}$ for the whole hypergraph~$G$. However, since a hyperedge~$e$ containing nodes~$i,j$ may be contained in one neighborhood backbone~$B^\ast_i$ but not another $B^\ast_j$, we must carefully consider how to combine these local backbones to form $B^{(\text{local})}$. This issue arises in the disparity filter methodology as well~\cite{serrano2009extracting}, where an edge may belong to the backbone neighborhood of one of its endpoints but pruned from the neighborhood of the other. Given that each hyperedge~$e\in G$ may have more than two nodes, there are a number of aggregation strategies for forming $B^{(\text{local})}$. 

One option is to just take the union of the neighborhood-level backbones, thus
\begin{align}\label{eq:union}
B_{\text{union}}^{(\text{local})} = \bigcup_{i=1}^{N}B^\ast_i.
\end{align}
This formulation has the advantage of guaranteeing that every node is included in the final backbone~$B^{(\text{local})}$, but in practice tends to give very dense backbones, since an edge may be in the backbone of only one of its constituent nodes which is the least well-connected in general. A better strategy is to take a ``majority rule''. Along with the union aggregation rule above, this is consistent with the disparity filter convention for pairwise edges in which the edge is kept in the full network backbone if it belongs to either node's neighborhood backbone. For the majority rule, we keep an edge~$e$ in $ B_{\text{majority}}^{(\text{local})}$ only if it appears in a majority of its constituent nodes' neighborhood backbones. Finally, the most conservative option is to include the edge~$e$ only if it is included in \emph{all} of its constituent node's backbones. However, this is often too restrictive in practice, resulting in very sparse hypergraph backbones~$B_{\text{all}}^{(\text{local})}$.

Given the final local backbone~$B^{(\text{local})}$, we have a total description length of
\begin{align}
\mathcal{L}^{(\textrm{local})}(G,B^{(\text{local})}) = \sum_{i=1}^{N}\mathcal{L}(G_i, B^{\ast}_i \cap B^{(\text{local})}),   
\end{align}
where the local neighborhoods are restricted to the edges that appeared in the final backbone. The corresponding compression ratio is then
\begin{align}
\eta^{(\text{local})} = \frac{\mathcal{L}^{(\text{local})}(G, B^{(\text{local})})}{\mathcal{L}^{(\text{local})}(G,G)}.    
\end{align}

\begin{figure*}[hb!]
\includegraphics[width=0.96\textwidth]{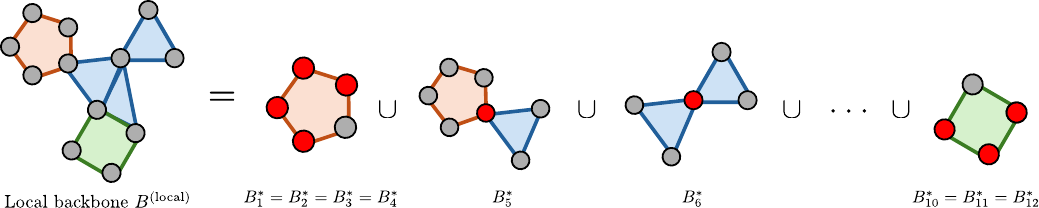}
\caption{
    \textbf{Local higher-order backboning.} 
    The local backboning method of Appendix~\ref{app:local-derivation} consists of applying the higher-order backboning procedure to each node neighborhood $G_i$ (the set of hyperedges containing node $i$) separately, then aggregating these neighborhood backbones. Here, the focal node $i$ for each neighborhood and its structurally equivalent nodes with identical neighborhood backbones are highlighted in red, and the union aggregation rule of Eq.~\eqref{eq:union} is applied.  
}
\label{fig:fig_Blocal}
\end{figure*}

This local backboning formulation has the advantage of being easily parallelizable across neighborhoods $G_i$ as well as accounting for heterogeneity in the connectivity of different nodes by considering the backbones only relative to each node neighborhood.

\begin{table*}[t!]
\centering
%\tiny
\rowcolors{2}{white}{gray!10}
\resizebox{0.5\textwidth}{!}{
\begin{tabular}{lcccccc}
    \hline\toprule\toprule
    \textbf{Dataset} & \multicolumn{1}{c}{$N$} & \multicolumn{1}{c}{$|G|$} & \multicolumn{1}{c}{$|B|$} & \multicolumn{1}{c}{$|B^{(\textrm{local})}|$} & \multicolumn{1}{c}{$\eta$} & \multicolumn{1}{c}{$\eta^{(\textrm{local})}$} \\
    \toprule\toprule\hline
    coauth-mag-geology\_1980 & 1674 & 903 & 310 & 810 & 0.80 & 0.88 \\
    coauth-mag-geology\_1981 & 1075 & 547 & 192 & 498 & 0.82 & 0.91 \\
    coauth-mag-geology\_1982 & 1878 & 987 & 345 & 854 & 0.80 & 0.88 \\
    coauth-mag-geology\_1983 & 1734 & 883 & 313 & 798 & 0.81 & 0.90 \\
    contact-high-school & 327 & 7818 & 2223 & 3142 & 0.65 & 0.68 \\
    contact-primary-school & 242 & 12704 & 3579 & 4799 & 0.70 & 0.72 \\
    dawn & 2290 & 138742 & 52704 & 108592 & 0.83 & 0.92 \\
    email-enron & 143 & 1459 & 309 & 420 & 0.61 & 0.67 \\
    email-eu & 986 & 24520 & 5484 & 8367 & 0.60 & 0.58 \\
    hospital-lyon & 75 & 1824 & 545 & 675 & 0.69 & 0.72 \\
    hypertext-conference & 113 & 2434 & 691 & 1096 & 0.74 & 0.76 \\
    invs13 & 92 & 787 & 222 & 81 & 0.77 & 0.74 \\
    invs15 & 217 & 4909 & 1415 & 2211 & 0.70 & 0.73 \\
    kaggle-whats-cooking & 6714 & 39224 & 8918 & 22712 & 0.78 & 0.87 \\
    malawi-village & 84 & 431 & 125 & 184 & 0.68 & 0.69 \\
    ndc-classes & 628 & 796 & 226 & 329 & 0.54 & 0.52 \\
    ndc-substances & 3414 & 6471 & 1549 & 2306 & 0.60 & 0.67 \\
    science-gallery & 410 & 3350 & 924 & 1392 & 0.64 & 0.65 \\
    sfhh-conference & 403 & 10541 & 2989 & 4495 & 0.70 & 0.71 \\
    tags-ask-ubuntu & 3021 & 145053 & 43770 & 91431 & 0.73 & 0.80 \\
    tags-math-sx & 1627 & 169259 & 57071 & 121253 & 0.75 & 0.83 \\
    \bottomrule
\hline
\end{tabular}
}
    \caption{Local higher-order backboning of real-world hypergraphs.
    }
    \label{tab:local}
\end{table*}

In Fig.~\ref{fig:fig_Blocal} we diagram the local backboning process for a small example network, using the union aggregation rule of Eq.~\eqref{eq:union}. We can see that this method has the benefit of identifying and removing structures that are redundant within each neighborhood separately, which controls for heterogeneity across these neighborhoods and more often produces well-connected backbones \cite{serrano2009extracting}. The downside of this is that the resulting backbones are often fairly dense in practice, which reduces the compression provided by the backbone.

In Table~\ref{tab:local} we repeat the analyses of Table~\ref{tab:datasets} using this local backboning procedure, using the majority rule for aggregation. We can see that, due to its purely local nature, the local method systematically produces more densely connected backbones $B^{(\text{local})}$ than the method of Sec.~\ref{sec:empirical} (denoted $B$ here). This comes at the price of worse compression in most cases, as shown by the values $\eta^{(\text{local})}$ in the rightmost column, which exceed the values of $\eta$ for the original method for all but a few networks.

\newpage

\section{Backboning real-world hypergraphs with weights}
\label{app:weighted-real}

In this section, we discuss the results of applying our backboning method to weighted real-world hypergraphs. To this end, we select a subset of
(originally unweighted) hypergraphs studied in Sec~\ref{sec:empirical} that are contact-based, as these are naturally represented as weighted hypergraphs with edge weights based on interaction intensity~\cite{del2026distances}. The datasets include contact-high-school, contact-primary-school, hospital-lyon, invs13, and sfhh-conference (see Table~\ref{tab:datasets} for more details on these datasets). For our analyses, we assigned the weight $w(e)$ to each hyperedge $e$ in the hypergraphs as the frequency of interactions observed over an appropriate time window $\Delta t$. 

\begin{figure*}[h!]
\includegraphics[width=1.00\textwidth]{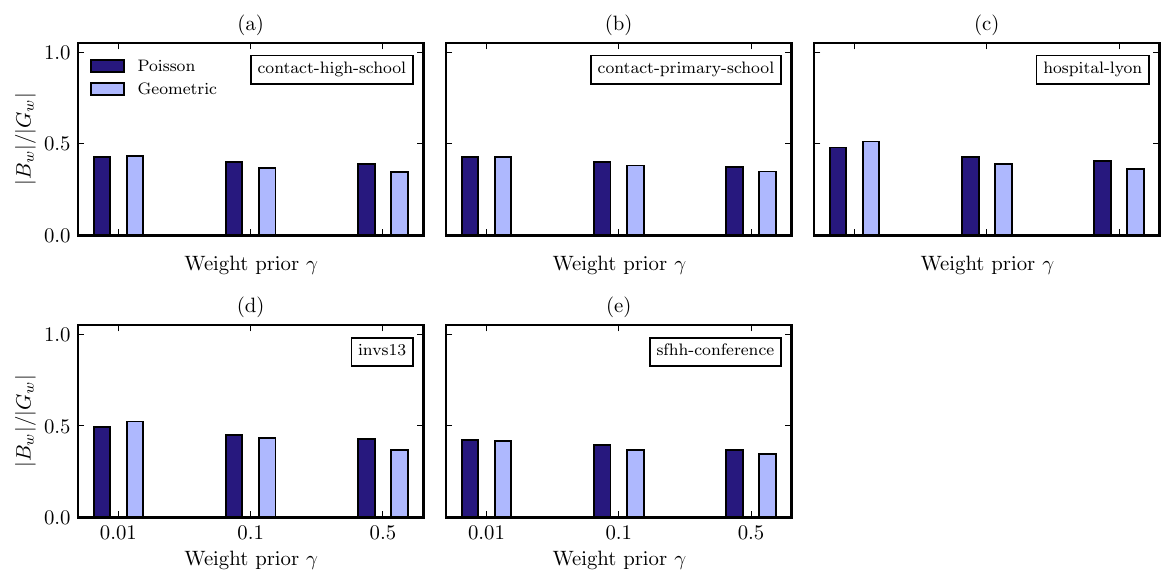}
\caption{
\textbf{Backboning weighted real-world hypergraphs.}
(a)--(e)~Fraction of hyperedges retained after applying the
weighted backboning method to the real-world contact datasets, for various values of $\gamma$ and different choices of weight prior.
}

\label{fig:figure_weighted_realworld}
\end{figure*}

Figure~\ref{fig:figure_weighted_realworld} shows the results of applying the backboning method of Sec.~\ref{sec:weighted} to these contact-based hypergraphs. For each dataset, we plot the fraction of hyperedges retained after applying the weighted backboning method, for different choices of the
weight prior and values of $\gamma$. Across all datasets, we observe a substantial reduction in the number of retained hyperedges. However, the fraction of edges retained in this weighted representation is generally higher than for the unweighted representation (in this case, $\sim 40-45\%$ of all hyperedges are kept in the backbone $B^\ast$). This is sensible because in the weighted representation, the backbones will tend to keep all hyperedges deemed topologically relevant, while also keeping additional hyperedges that can be justifiably placed in the backbone due to their high weight. 

The extent to which hyperedges will be kept in the backbone purely due to weight will in principle depend on the parameter $\gamma$ (see Sec.~\ref{sec:weighted}), and indeed we see a drop in the fraction of retained hyperedges as $\gamma$ increases as expected. However, the results are fairly insensitive to $\gamma$ for most datasets and priors studied here, indicating that the exact choice of this hyperparameter is not highly consequential in this case. We also see very little variation in the results as the weight prior changes, indicating that this choice is also not essential for performance of the method. 

\newpage

\section{Comparison with structural reducibility}
\label{app:reducibility}

\begin{figure*}[h!]
\includegraphics[width=1.0\textwidth]{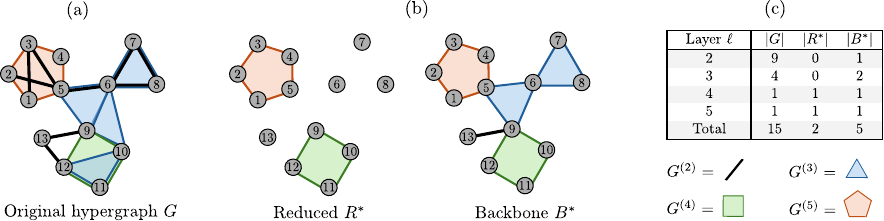}
\caption{
    \textbf{Structural reducibility and hypergraph backboning.}
    (a)~Example hypergraph~$G$ from Fig.~\ref{fig:fig1_diagram}.
    (b)~Reduced hypergraph~$R^\ast$, obtained using the method
    in~\cite{kirkley2025structural}, and structural backbone~$B^\ast$.
    (c)~Number of hyperedges of each order $\ell\in \{2,3,4,5\}$ in
    $G$, $R^\ast$, and $B^\ast$. The reduced hypergraph globally
    removes entire redundant interaction orders (here, discarding all
    dyads and triplets) without accounting for local structural
    variations. Meanwhile, the backbone~$B^\ast$ accounts for local heterogeneities and preserves
    structurally non-redundant interactions across all orders.
}
\label{fig:fig:fig1_supp-diagram}
\end{figure*}

The backboning framework developed in this paper is somewhat similar in spirit to, but conceptually distinct from, the notion of \emph{reducibility}
of higher-order networks~\cite{lucas2026reducibility,kirkley2025structural}. While reducibility uses aims to compress a
hypergraph, it operates at the level of entire interaction orders (hypergraph layers) by
identifying the orders that most parsimoniously summarize the system
and retains \emph{all} hyperedges of the selected orders while
discarding \emph{all} hyperedges of the remaining ones, yielding a
reduced hypergraph $R^\ast$ (see Fig.~\ref{fig:fig:fig1_supp-diagram}).

The crucial distinction between methods is therefore one of granularity and
locality. Structural reducibility makes a single, global decision per order, implicitly assuming that each
order contributes homogeneously to the structure across the entire
hypergraph. Our backboning method instead selects a subset $B\subseteq
G$ of individual hyperedges by exploiting pairwise nestedness and
overlap through a parent-child encoding, so that the effective set of
retained orders is free to vary from one region of the hypergraph to
another. This flexibility is what allows the backbone to preserve structurally central low-order interactions while pruning locally redundant high-order
interactions. This situation is illustrated in
Fig.~\ref{fig:fig:fig1_supp-diagram}, where order-level reduction disconnects
part of the system while the backbone retains its connectivity. In
this sense, structural reducibility can be viewed as a constrained
special case of the more general sparsification problem that
backboning solves, in which the retained set is forced to be a union of complete layers.

\begin{figure}[h!]
\includegraphics[width=1.0\textwidth]{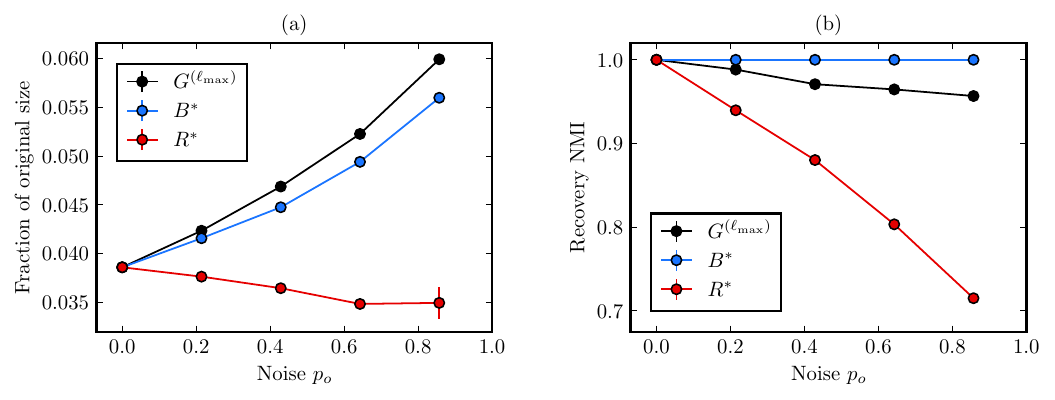}
\caption{
    \textbf{Reducibility and backboning for synthetic hypergraphs with tunable structure.}
    (a)~Fraction of edges retained in the backbone hypergraph $B^\ast$ inferred using Eq.~\eqref{eq:Ltotalorig} as well as the representation $R^\ast$ inferred by applying the structural reducibility method in \cite{kirkley2025structural}, for the noise $p_o$ of the order sensitivity test described in this Appendix.
    (b)~Similarity, measured using the hypergraph NMI of \cite{felippe2026information}, between the top faces of the noisy simplices and $B^\ast,R^\ast$. 
    As a baseline, we show the recovery accuracy obtained when only keeping the top layer $G^{(\ell_{\text{max}})}$ of the resulting hypergraph. 
}
\label{fig:fig2supp}
\end{figure}

We further investigate the practical consequences of this distinction in 
synthetic data with controlled interaction orders, in which we generate fully nested
hypergraphs (i.e., simplicial complexes) on $100$ nodes by sampling
$10$ simplices, each assigned a top face of size $5$ with probability
$1-p_o$ and a top face whose size is drawn uniformly from
$\{2,3,4\}$ with probability $p_o$, so that increasing $p_o$ makes the
dominant interaction orders more heterogeneous (results averaged over
$100$ realizations, with error bars indicating $3$ standard errors in
the mean). We compare the
inferred hypergraph backbone $B^\ast$ against the reduced representation $R^\ast$
obtained with the structural reducibility method
of~\cite{kirkley2025structural} by applying both methods to these synthetic hypergraphs. In Fig.~\ref{fig:fig2supp}(a) we plot
the fraction of edges retained by each method: the backbone
consistently recovers the correct planted size regardless of which
orders are present, whereas reducibility degrades rapidly as the
dominant orders become more diverse, since it must keep or discard
entire orders at once. In Fig.~\ref{fig:fig2supp}(b) we plot the
hypergraph NMI~\cite{felippe2026information} between the true top faces
of the generated simplices and the two representations. The backbone
recovers the dominant hyperedges with near-perfect accuracy across all
levels of size heterogeneity, while reducibility fails to reflect the
underlying nested structure---so much so that a naive baseline that
simply keeps the largest layer $G^{(\ell_{\text{max}})}$ substantially
outperforms it. Together these results confirm that the additional
locality afforded by hyperedge-level selection translates into markedly
better structural recovery whenever the relevant interaction orders
vary across the system.

\end{document}